\documentclass[letter]{aa} 
\usepackage{amsmath}
\usepackage{natbib}
\usepackage{graphicx}
\usepackage{txfonts}
\usepackage{longtable}
\newcommand{\kmso}{\hbox{${\rm km}\:{\rm s}^{-1}$}}
\newcommand{\teff}{$T_{\rm eff}\;$}  
\newcommand{\teffo}{$T_{\rm eff}$}

\bibpunct{(}{)}{;}{a}{}{,} 
\begin{document}
\title{Lithium in the globular cluster NGC 6397\thanks{Based
on observations obtained with FLAMES/GIRAFFE at VLT Kueyen 8.2 m
telescope in programme 079.D-0399(A)}\subtitle{Evidence for dependence on evolutionary status.}}
\titlerunning{Cosmological Li problem unsolved.}
\authorrunning{Gonz\'alez Hern\'andez et al.}
%
%
\author{J.~I.~Gonz\'alez Hern\'andez\inst{1,2}\and 
P.~Bonifacio\inst{1,2,3}\and 
E.~Caffau\inst{1}\and M.~Steffen\inst{4}\and 
H.-G.~Ludwig\inst{1,2}\and 
N.~T.~Behara\inst{1,2}\and L.~Sbordone\inst{1,2}\and
R.~Cayrel\inst{1} \and S.~Zaggia\inst{5}}
\offprints{J.~I. Gonz\'alez Hern\'andez.}
\institute{
Cosmological Impact of the First STars (CIFIST) Marie Curie Excellence Team
\and
GEPI, Observatoire de Paris, CNRS, Universit\'e Paris Diderot; Place
Jules Janssen 92190
Meudon, France \\
\email{Jonay.Gonzalez-Hernandez@obspm.fr}
\and
Istituto Nazionale di Astrofisica - Osservatorio Astronomico di
Trieste, Via Tiepolo 11, I-34143  Trieste, Italy
\and
Astrophysikalisches Institut Potsdam, An der Sternwarte 16, D-14482
Potsdam, Germany 
\and
INAF - Osservatorio Astronomico di Padova, Vicolo dell'Osservatorio 5,
Padua 35122, Italy} 

\date{Received 16 June 2009; accepted 4 September 2009}
 
\abstract
{
Most globular clusters are believed to host a single stellar
population. They can thus be considered a good place to study the
Spite plateau and to search for possible evolutionary modifications
of the Li content.
}  
{We want to determine the Li content of subgiant (SG) and
main sequence (MS) stars of the old, metal-poor
globular cluster NGC 6397. This work was aimed not only at studying
possible Li abundance variations but also to investigate the cosmological
Li discrepancy.}
{Here, we present FLAMES/GIRAFFE observations of a sample of 84 SG and
79 MS stars in NGC 6397 selected in a narrow range of $B-V$ colour
and, therefore, effective temperatures. We determine both effective
temperatures and Li abundances using three-dimensional hydrodynamical
model atmospheres for all the MS and SG stars of the sample.
}    
{We find a significant difference in the Li abundance between SG stars
and MS stars, the SG stars having an abundance higher by almost
0.1\,dex on average. We also find a decrease in the lithium abundance
with decreasing effective temperature, both in MS and SG stars, albeit
with a significantly different slope for the two classes of stars.  
This suggests that the lithium abundance in these stars is, indeed,
altered by some process, which is temperature-dependent.
}   
{The lithium abundance pattern observed in NGC 6397 is different from
what is found among field stars, casting some doubt on the use of
globular cluster stars as representative of Population II with respect
to the lithium abundance. None of the available theories of Li
depletion appears to satisfactorily describe our observations. 
}   

\keywords{Stars: abundances -- Stars: atmospheres
-- Stars: fundamental parameters -- Stars: Population II -
(Galaxy:) globular clusters: individual: NGC 6397}

\maketitle

\section{Introduction\label{intro}}

The old, metal-poor dwarf stars of the Galactic halo share
approximately the same Li abundance, irrespective of their metallicity
or effective temperature \citep{spi82a,spi82b}. This plateau
of lithium was believed to provide evidence of a primordial Li
abundance. 
The WMAP satellite has been able to measure with high accuracy the
baryonic density from the fluctuations of the cosmic
microwave background \citep{spe07}. This result implies a primordial Li
abundance of $\log ({\rm Li}/{\rm H})+ 12 =2.72\pm0.06$ \citep{cyburt}
whereas the observed Li abundances in metal-poor dwarfs are in the
range 2.0-2.4 
\citep[see][and references therein]{sbor08,bon07,asp06,cap05,mel04}.  
This discrepancy may be trivially solved if the {\em Spite plateau}
does not represent the primordial Li abundance.  
In this case the amount of lithium in the atmospheres of all ancient
stars, of all masses and metallicities, must have been uniformly
depleted by at least a factor of three. 

Possible explanations of this difference are: (a) the first generation
of stars, Population III stars, could have processed some fraction of
the halo gas, lowering the lithium abundance \citep{pia06}; 
(b) the primordial Li abundance has been uniformly depleted in the
atmospheres of metal-poor dwarfs by some physical mechanism (e.g.
turbulent diffusion as in \citealt{ric05,kor06}; gravitational
waves as in \citealt{cat05}, etc.); (c) the standard Big Bang
nucleosynthesis (SBBN) calculations should be revised, possibly with
the introduction of new physics 
\citep[see e.g.][]{jed04,jed06,jittoh,hisano}. 
The observed Li abundances, A(Li), in metal-poor stars appear
to show a very well defined plateau with very little dispersion at
relatively high metallicities, whereas at low metallicities there
seems to be an increased scatter, or perhaps even a sharp down turn
in the Li abundances \citep{bon07,gon08,sbor08}.
The existence of a slope in A(Li) versus [Fe/H] would exacerbate the
discrepancy between Li abundance in metal-poor stars and the WMAP
predictions \citep{bon07}. The issue of the slope
in the plateau is somewhat elusive and different groups reach
different conclusions, depending on the adopted temperature
scale \citep{mel04}. 

Globular clusters (GCs) were initially considered to be a 
good place to investigate the Spite plateau
\citep{MP94,PM96,PM97,boe98}, since the classical
paradigm was that GCs are made of a single stellar population.
The discovery of correlations among elemental abundances in turn-off (TO)
stars \citep{gratton} and in particular the presence
of a Li-Na anti-correlation \citep{pasquini,bonifacio},
showed the need for the presence of different
stellar populations, capable of nucleosynthetic
activity and variable amounts of pollution 
of the presently observable stars.
Such signatures are not found among field stars
and are peculiar to GCs.
This makes the perspective of using GCs to investigate
the Spite plateau meagre.

Among the observed GCs, NGC 6397 occupies a special
role, in the sense that the Li abundance among
non-evolved stars is very homogeneous \citep{theve,bon02}, 
at variance with what is observed in NGC 6752
\citep{pasquini} and 47 Tuc \citep{bonifacio}.
More recent studies \citep{kor06,kor07} have claimed
a tiny variation of A(Li) along
the subgiant branch, in the sense of
higher A(Li) being found for lower \teff values. 
This variation is however quite small compared
to what is observed in NGC 6752 and 47 Tuc.

In this Letter we present the result of the 
analysis of the first observations of Li 
in main sequence (MS) stars of a globular cluster.

\begin{figure}
\centering
\includegraphics[width=8.5cm,angle=0.]{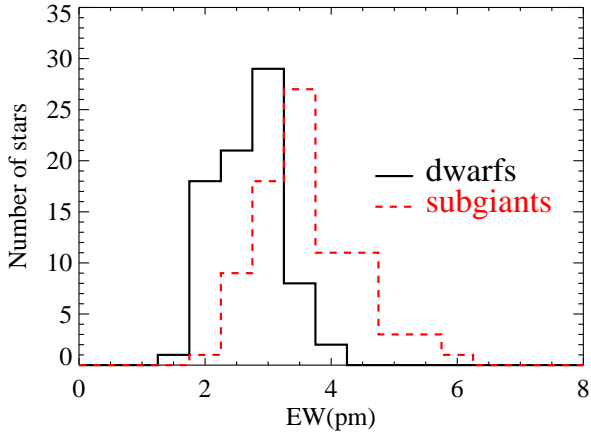}
\caption{\footnotesize{Histograms of observed equivalent width
of the lithium doublet at 670.8~nm in MS and SG stars in the
globular cluster NGC 6397. Histograms of the equivalent width of Li
line are displayed in bins of 0.5~pm for MS stars (solid line) and
SG stars (dashed-dotted line).}} 
\label{figew}   
\end{figure}
  
\section{Observations}

\begin{figure}
\centering
\includegraphics[width=9.0cm,angle=0.]{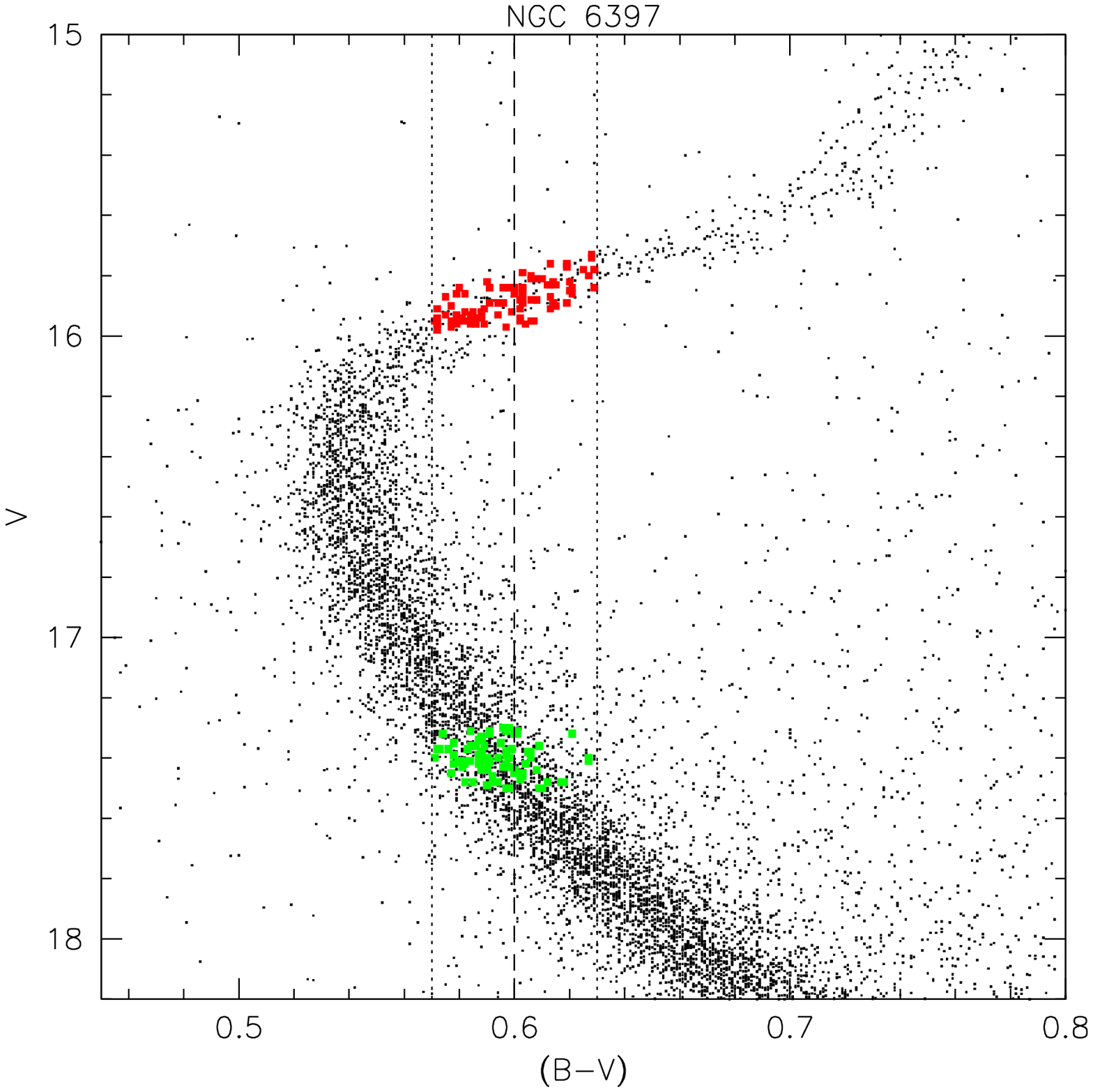}
\caption{\footnotesize{Colour-magnitude diagram of the cluster NGC
6397. The stars studied in this work are depicted in small filled
squares.}}
\label{figcmd}   
\end{figure}

\begin{figure}
\centering
\includegraphics[width=8.5cm,angle=0]{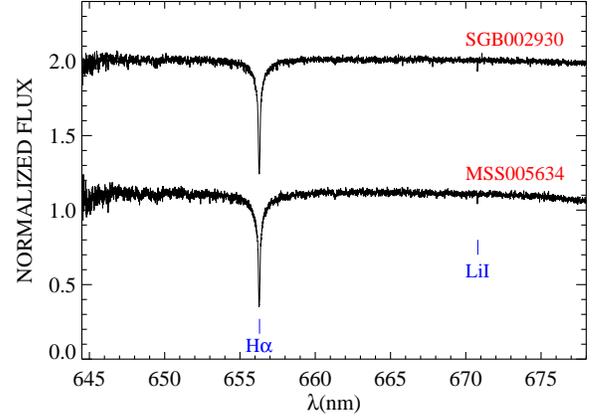}
\caption{\footnotesize{Observed GIRAFFE/FLAMES spectra of a dwarf star
MSS005634 (bottom, S/N~$= 102$) and a subgiant star SGB002930 (top,
S/N~$= 111$)of the globular cluster NGC~6397.}} 
\label{figspec}   
\end{figure}

We integrated NGC 6397 over 15 hours   
with the multi-object spectrograph FLAMES-GIRAFFE \citep{pas02} at
the European Southern Observatory (ESO), using the 8.2-m Very Large
Telescope, on 2007 April-July, covering the spectral
range $\lambda\lambda$6400--6800 {\AA} at a resolving power
$\lambda/\delta\lambda\sim17,000$.
The targets were selected using our own calibrated Johnson-Cousins
B, V photometry, based on public images (ESO program 163.O-0741(C))
obtained with WFI at the ESO/MPI 2.2m telescope on 14 May 1999.
We chose SG and MS stars in the colour range
$B-V=0.6\pm0.03$, thus ensuring a narrow \teff range
(see Fig.~\ref{figcmd} online).
By swapping the fibres on the SGs we managed to observe over 9 hours
for about 80 MS stars and 2.5 hours for roughly the same number of SGs. 
The resulting S/N ratio $\sim80-130$ is the same for both sets of
stars. The spectra were reduced using the ESO Giraffe pipeline,
version 2.5.3.  
A combined spectrum of all sky fiber spectra in each night was
properly subtracted from each individual spectrum. We then corrected
each spectrum for the earth velocity and combined all the spectra
of the same target (see the quality of the spectra in
Fig.~\ref{figspec} online).
Each star spectrum was corrected for its radial velocity, providing a
mean cluster radial velocity of $V_{r,c}=18.5$~\kmso.
We removed all stars, considered as cluster non-members,
with $\mid V_r-V_{r,c} \mid~> 3\sigma_{V_{r,c}}$, where
$\sigma_{V_{r,c}}$ is the radial velocity dispersion (3.7~\kmso). We
ended up with 79 MS (orginally 80) and 84 SG (88). 

\section{Analysis and results}
 
\begin{figure*}
\centering
\includegraphics[width=8.5cm,angle=0.]{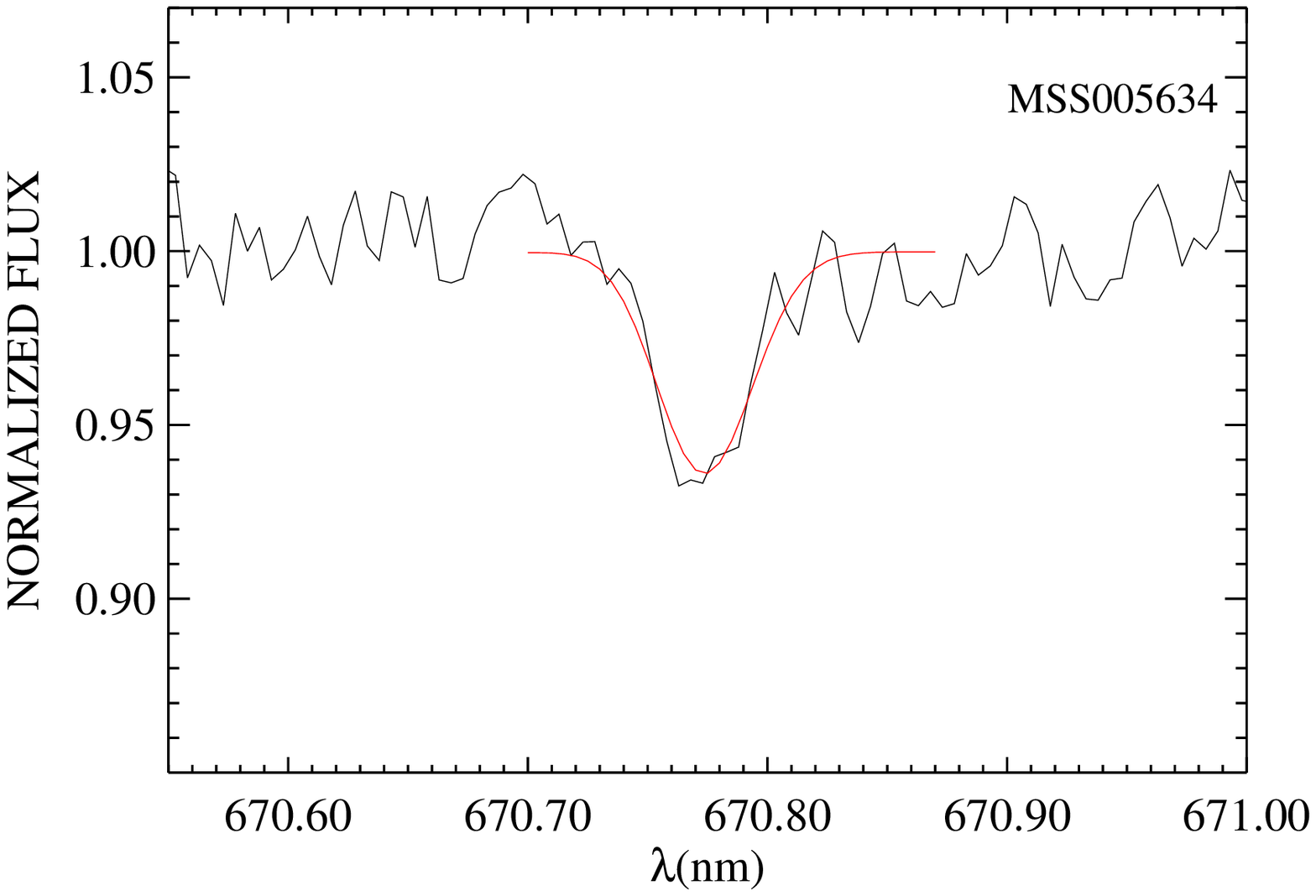}
\includegraphics[width=8.5cm,angle=0.]{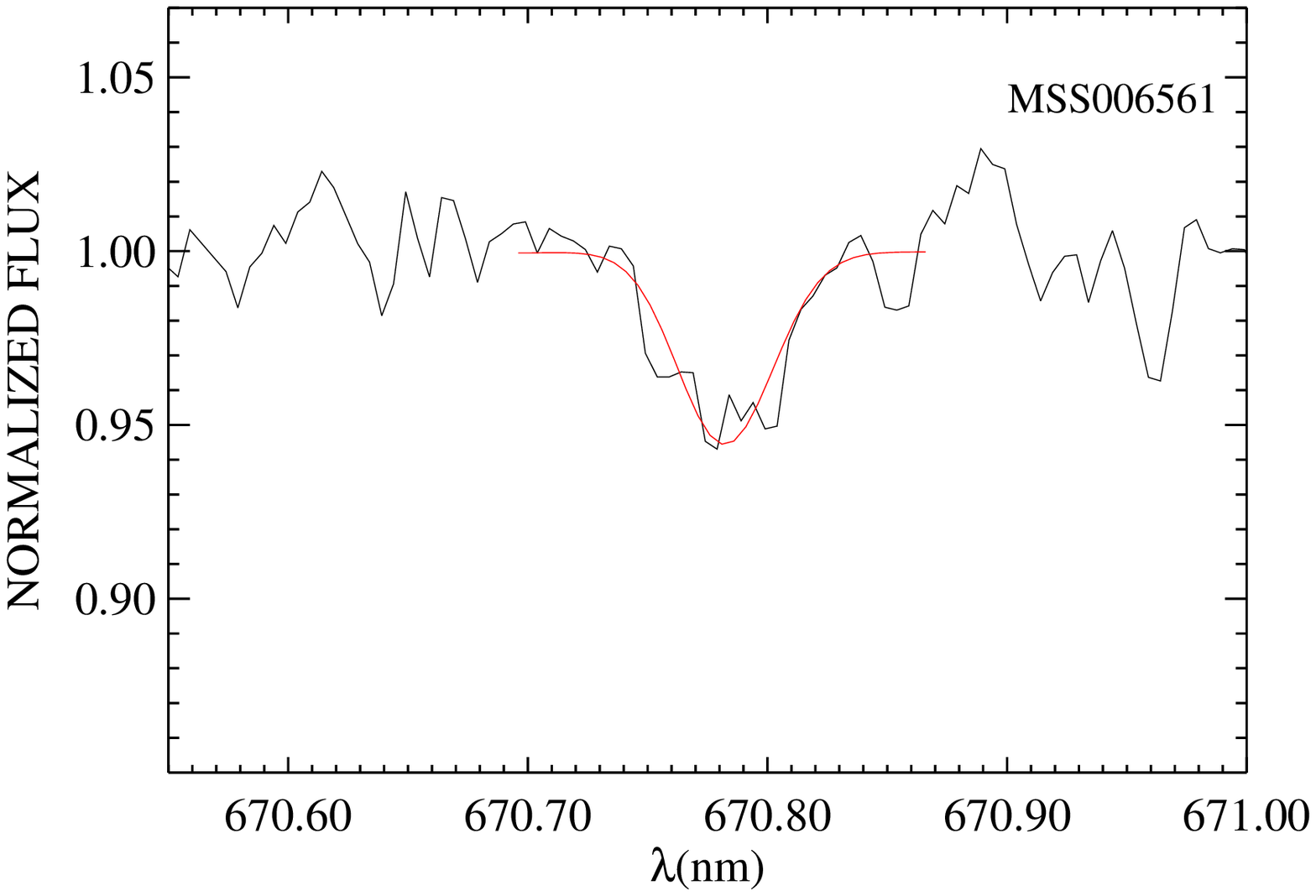}
\includegraphics[width=8.5cm,angle=0.]{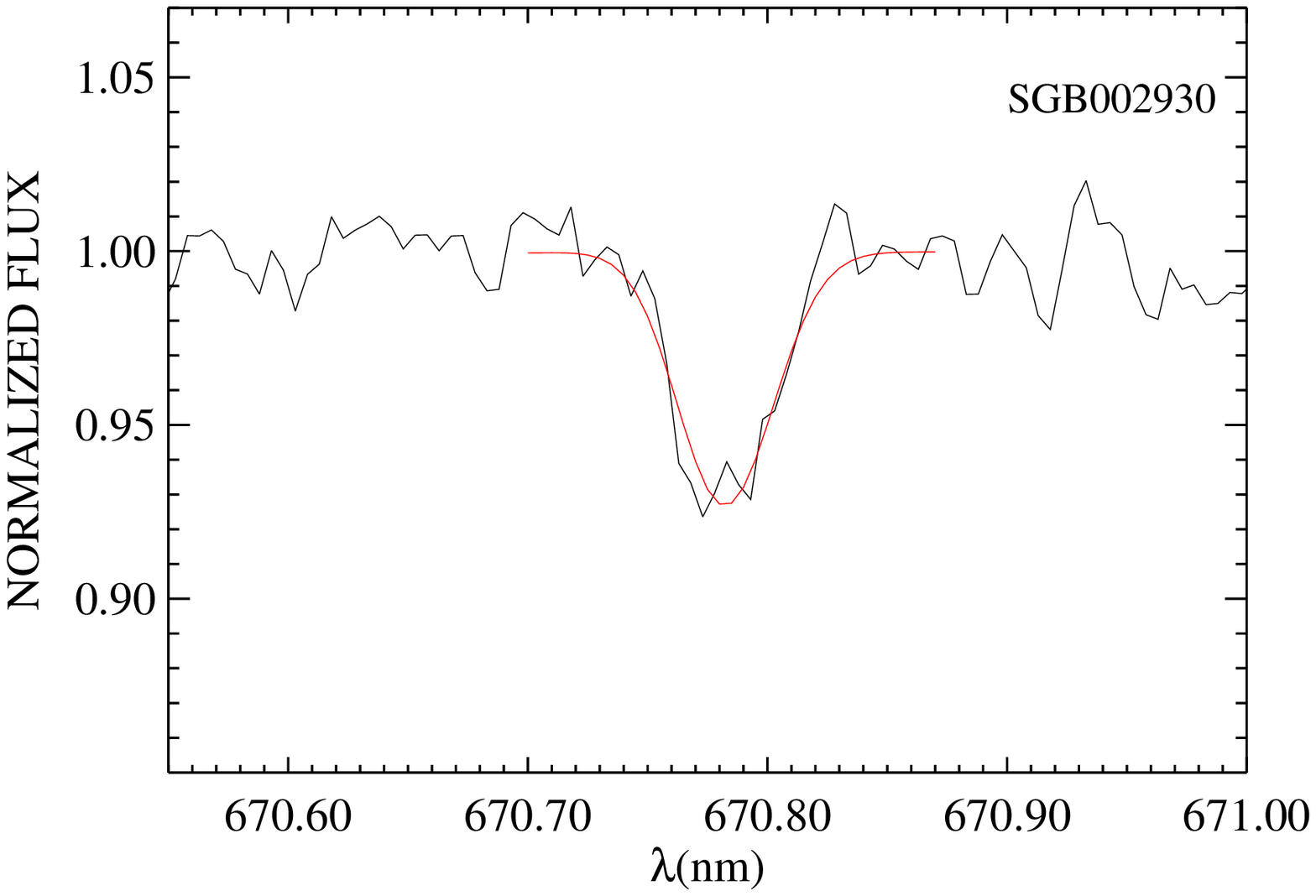}
\includegraphics[width=8.5cm,angle=0.]{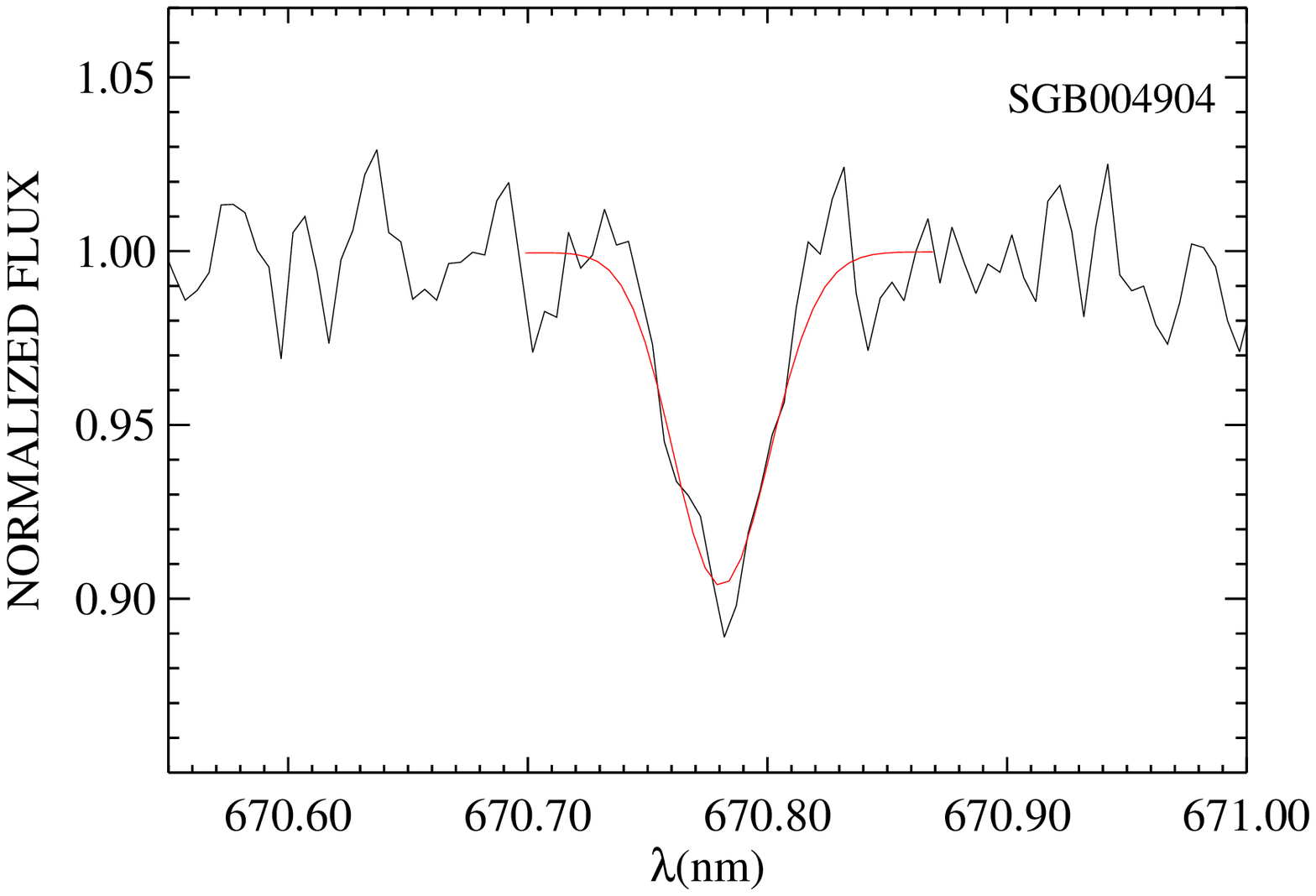}
\caption{\footnotesize{Observed spectra of two dwarf stars, MSS005634
(top-left panel, S/N~$= 102$, EW(Li)~$=32.21$~m{\AA}) and MSS006561 
(top-right panel, S/N~$=71$, EW(Li)~$=27.97$~m{\AA}) and two subgiant
stars, SGB002930 (bottom-left panel, S/N~$= 111$,
EW(Li)~$=36.83$~m{\AA}) and SGB004904 (bottom-left panel, S/N~$=68$,
EW(Li)~$=48.52$~m{\AA}), showing the fit of the Li line with a
synthetic profile.}}  
\label{figewobs}
\end{figure*}

The narrow range in effective temperatures  
ensures that the uncertainty in the comparisons between MS and
SG stars is dominated by the error on the measured equivalent width
(EW) of the Li doublet line.
These were measured by fitting the observed Li line profile with
synthetic profiles of the Li doublet, as previously done for this
cluster \citep{bon02}. The EW measurements show 
(Fig.~\ref{figew}) that SGs have, on average, larger
EWs of the Li doublet than MS stars (see the
accuracy of our fitting procedure in Fig.~\ref{figewobs} online). 
Although there
is a slight dependence of the $B-V$ colour on surface gravity, and a
SG star of a given colour is indeed cooler than a MS star of
the same colour ($\sim 90$~K at $B-V=0.6$), 
the difference displayed in Fig.~\ref{figew} is too
large to be explained in this way. 
The weighted mean of the EW is $2.97\pm0.02$~pm and
$4.06\pm0.01$~pm for MSs and SGs, respectively.
The difference in the mean EW values is of about 1.1~pm
which would require a mean \teff difference of $\sim 210$~K.
Prior to any model-dependent analysis, this clearly points 
towards the SGs having a higher Li abundance than the MSs.
This is similar to what is found among field stars, where the Li
abundance appears to be about 0.04\,dex higher in turn-off and
SG stars than in MSs \citep{cap05}. 

\begin{figure*}
\centering
\includegraphics[width=8.5cm,angle=0.]{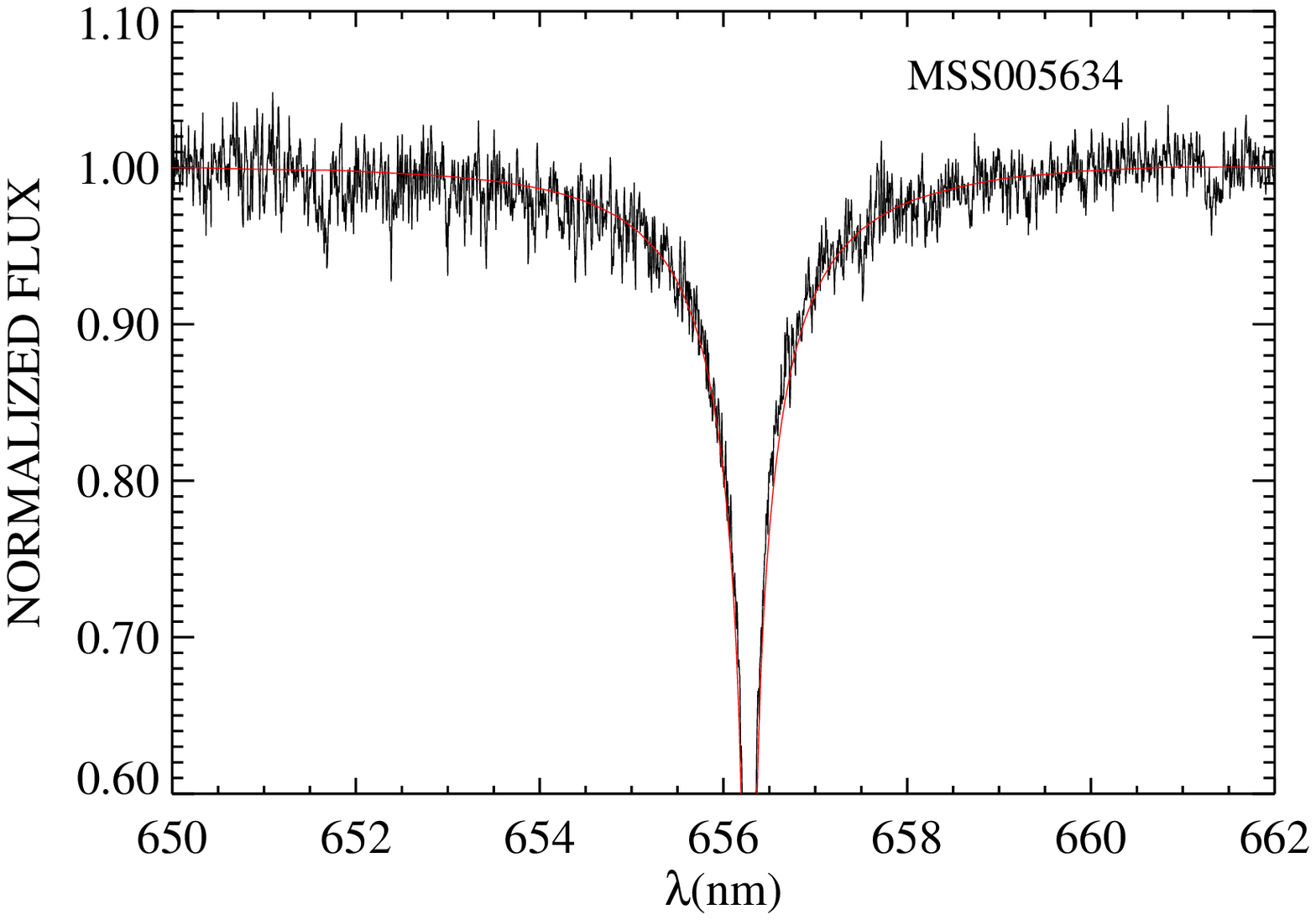}
\includegraphics[width=8.5cm,angle=0.]{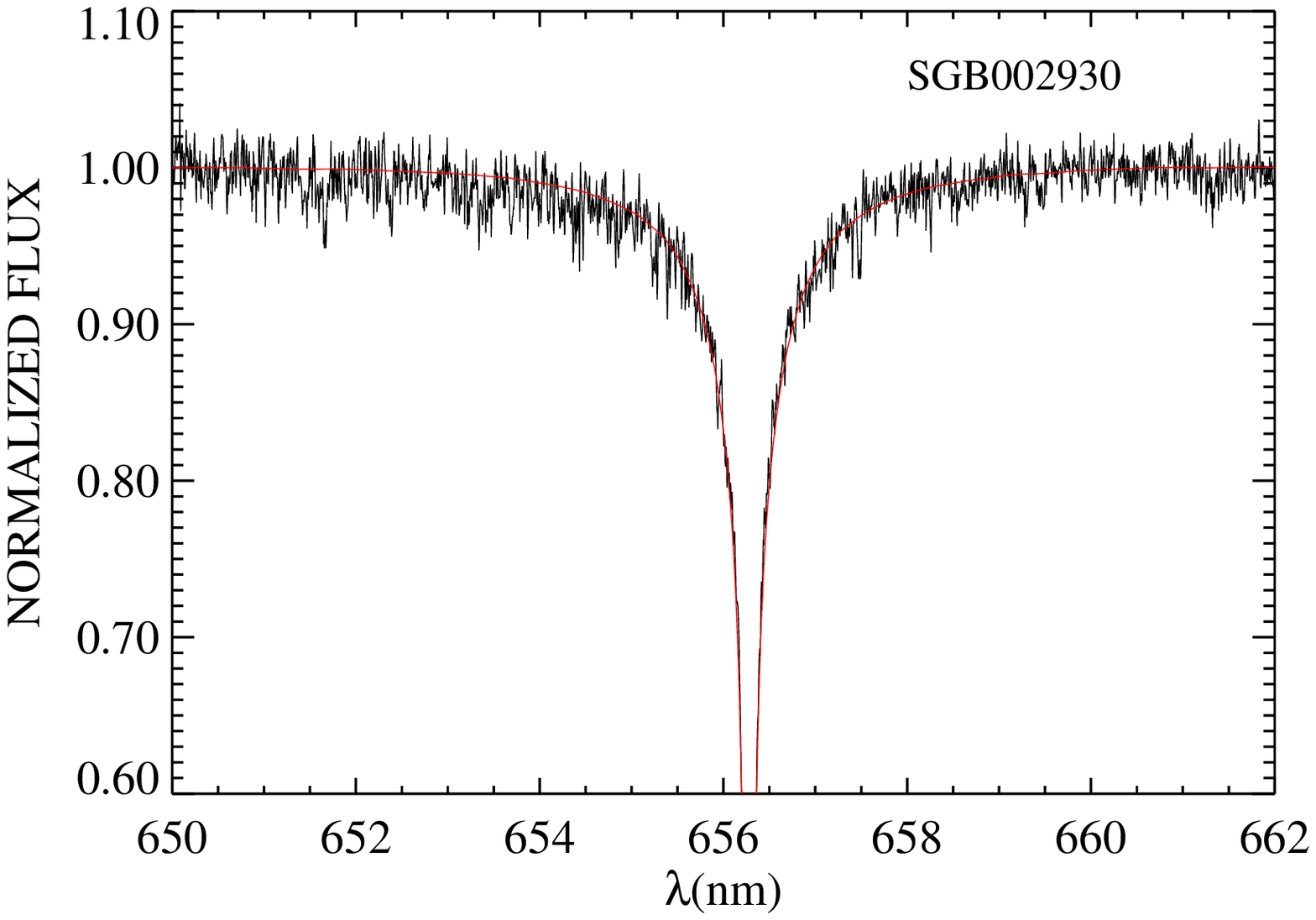}
\caption{\footnotesize{Observed GIRAFFE/FLAMES H$\alpha$ profile
fitted with a synthetic 3D profile for a dwarf star MSS005634 (left
panel, S/N~$= 102$, $T_{\rm eff,3D}=6327$~K) and for a subgiant star
SGB002930 (right panel, S/N~$= 111$, $T_{\rm eff,3D}=6126$~K).}} 
\label{fighalpha}   
\end{figure*}

We derived \teff by fitting the observed
H$\alpha$ line profile with synthetic profiles, using 3D
hydrodynamical model atmospheres computed with the CO$^5$BOLD code
\citep{fre02,wed04}. The ability of 3D models to reproduce Balmer line
profiles has been shown in \citet{beh09}, where the H$\alpha$ profiles
of the Sun, and the metal-poor stars HD 84937, HD 74000 and HD 140283
were investigated. From a purely theoretical point of view
\citet{lud09} quantified the differences in using 1D or 3D
models for Balmer line fitting (see the accuracy of our fitting
procedure in Fig.~\ref{fighalpha} online). 
In the online Table~\ref{tabmod} we 
provide information on the 3D model atmospheres used in this work.
Self-broadening of the H$\alpha$ line was calculated according to
Barklem's theory \citep{bar00}. Stark broadening was
calculated following Griem's theories \citep{gri60} with 
corrections to approximate the Vidal et al. profiles \citep{vid73}.
Fixed values for the surface gravity were adopted for both SG
and MS stars in the sample, according to the values that best match
the position of the stars on a 12 Gyr isochrone \citep{str97}. 
The adopted values were $\log (g/{\rm cm~s}^2)=4.40$ and 3.85 for MSs
and SGs, respectively.  
This choice of the surface gravity is supported by the 1.6 magnitude
difference in the $V$-filter between SGs and MSs in the
sample. The Li abundances were derived using the
same 3D hydrodynamical model atmospheres. 
The line formation of Li was treated in non-local thermodynamical
equilibrium (NLTE) using the same code and model atom used in
\citet{cay07}. The model atom consists
of 8 energy levels and 11 transitions. Full details
will be given in Sbordone et al. (in preparation). To derive 3D-NLTE
Li abundances we used the analytical fit as a function of stellar
parameters and EW also provided in Sbordone et al. (in preparation).
The analysis was also done with 1D model
atmospheres, providing essentially the same picture, although
\teff in 1D show lower values.
We also tried using the \citet{carlsson} NLTE corrections,
rather than our own, with no significant difference in the
general picture.

\begin{figure}
\centering
\includegraphics[width=9.cm,angle=0]{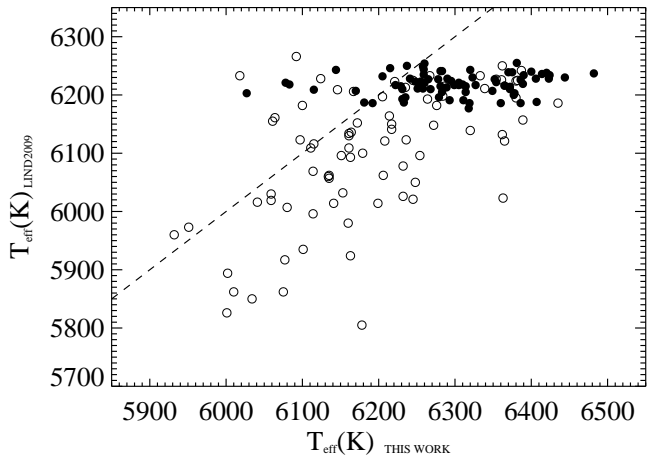}
\caption{\footnotesize{Comparison between 3D effective
temperatures of the observed stars and the 1D effective temperatures
derived from colors by Lind et al. (2009). Filled circles and
open circles correspond to dwarf and subgiant stars, respectively. The
dashed line shows the one-to-one relationship. Since our stars
have been selected in a $B-V$ range of 0.06 mag, their temperature
range should be of, at least 250 K. It could be larger due
to stars being moved into our selection box by photometric and reddening
uncertainties. There is no plausible reason why this range should 
be as small as that implied by the Lind et al. (2009) effective
temperatures ($\sim 80$K.}} 
\label{figteffteffl}
\end{figure}

\begin{figure}
\centering
\includegraphics[width=9.cm,angle=0]{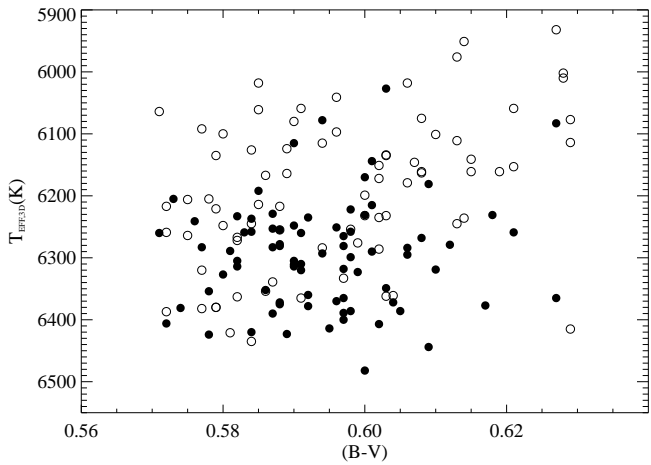}
\caption{\footnotesize{Comparison between 3D effective
temperatures and $B-V$ colours of the observed stars. 
Filled circles and open circles correspond to 
dwarf and subgiant stars, respectively. The lack of correlation
between $B-V$ and effective temperature is consistent with 
photometric errors and reddening variations.}} 
\label{figteffbv}
\end{figure}

In Fig.~\ref{figali} we display the derived Li 
abundances versus the effective temperatures of MSs and 
SGs of the globular cluster NGC 6397. 
The Li abundance decreases with decreasing temperature,
although more rapidly for MSs than for SGs.
This Li abundance pattern is different from what is
found among field stars \citep{mel04,bon07,gon08}.
The lithium-temperature correlations have a probability of 99.9\% and
99.5\% for MSs and SGs, respectively,
according to the non parametric rank correlation test, Kendall's
$\tau$ test. We performed a Kolmogorov-Smirnov test and obtained
a probability of $8\times10^{-6}$. Therefore, the possibility that the
two sets (MSs and SGs) have been drawn from the same
population (same Li abundance) can be rejected. 
Even ignoring the trend in A(Li) one can deduce that there 
is a real difference in the A(Li) of MSs and SGs by
computing the mean A(Li) and the standard deviation of the mean for
the two samples. For SGs we find $2.37\pm0.01$, while for MSs
$2.30\pm0.01$. 
Such a result is also evident in the analysis of
\citet{lin09} who find only a 0.03~dex difference
between the mean A(Li) in MSs and SGs, which is still
significant at 1$\sigma$. The signal is partly erased  
by the very narrow range of \teff for MSs deduced by \citet{lin09}
($\sim 80$~K) compared to the wide range 
($\sim 450$~K) for the SGs (see Fig.~\ref{figteffteffl}). 
Such a difference in the \teff range spanned by MSs and SGs 
is inconsistent with the very similar $B-V$ colours of 
the two sets of stars. In Fig.~\ref{figteffbv} online, the lack of
correlation between colour and \teff is fully compatible with 
the photometric and reddening uncertainties. 
The \teff values adopted by \citet{lin09} for the MSs are on 
the lower \teff side of the range spanned by the sample; 
this results in an artificial increase 
of the deduced A(Li) for the MSs,
which reduces the difference with SGs, without totally erasing it.
We conjecture that this is because the \teff estimates
of \citet{lin09} are derived by interpolating our $V$ magnitudes
onto the cluster fiducial sequence, ignoring any colour information. 
This necessarily compresses the \teff scale into a range smaller
than what is implied by the range in colour, when photometric errors and
variations in reddening are taken into account.

\section{Discussion and conclusions}     

\begin{figure}
\centering
\includegraphics[width=8.5cm,angle=0]{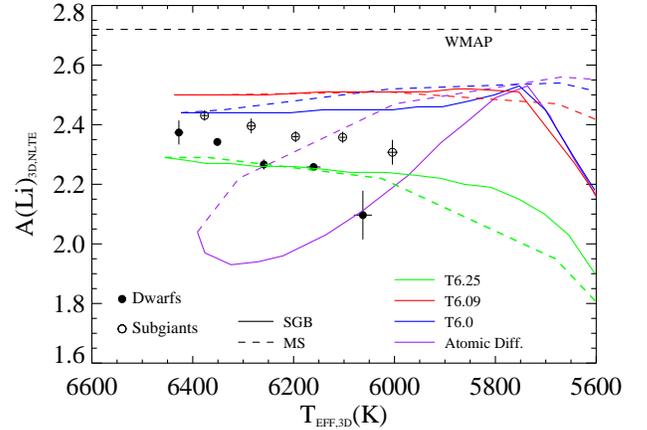}
\caption{\footnotesize{3D NLTE Li abundances versus 3D effective
temperatures of the observed MS (filled circles) and SG (open circles)
stars together with Li isochrones for different turbulent diffusion
models. 
The stars have been divided into five effective temperature bins. The
error bar in A(Li) shows the dispersion divided by the square root of
the number of stars in each bin. In each isochrone, the
dashed and solid stretch of the line shows the Li abundance in MS and
SG stars, respectively. The horizontal dashed line depicts the
cosmological Li abundance.}} 
\label{figali} 
\end{figure}

Our results imply unambiguously that the Li surface abundance changes
with evolutionary status. The fact that A(Li) is higher in SG stars
suggests a scenario in which lithium sinks below the photosphere during the
MS phase, but to a depth low enough to prevent Li distruction, so that
it can be restored in the photosphere, when the stars evolve beyond
the TO. The slope of A(Li) with \teff among MS
stars suggests that the amount by which Li is depleted in the
atmospheres is different for stars of different mass (\teff on the
MS). The similar slope found among SG stars suggests that 
after being restored in the atmosphere at the TO, 
lithium is then decreased by some other mechanism,
possibly mixing linked to the convective motions
which are more pronounced for the cooler \teff
of the SGs.
Although the above described scenario is plausible, 
we have so far no detailed understanding of
the physical processes that bring it about.
Diffusive processes  may alter the elemental composition of stars. 
Diffusion has been studied for decades \citep{aac60,mic84}, 
but only a few years ago, detailed element-by-element predictions 
from models including effects of atomic diffusion and 
radiative accelerations have become available \citep{ric02}.
These models produced strong abundance trends that are not
compatible with the Spite plateau, and only with the recent
inclusion of turbulent mixing, some of the model predictions roughly 
agree with observations \citep{ric05}.

Pure diffusion models \citep{ric05}, with no turbulence, predict
A(Li) differences as large as 0.4 dex between MSs and
SGs of the same age and temperature. 
The inclusion of turbulence can change this trend, and the SGs
may exhibit a A(Li) which is higher, lower, or almost
equal to that of the MSs, depending on the precise value of the
turbulence parameter.  

In Fig.~\ref{figali} we show the Li isochrones for different turbulent
diffusion models \citep{ric05}. These models have been shifted up by
0.14 dex in Li abundance to make the initial abundance of the models,
$\log ({\rm Li}/{\rm H})=2.58$, coincide with the primordial Li
abundance predicted from fluctuations of the microwave background
measured by the WMAP satellite \citep{cyburt}. 

The models assuming pure atomic diffusion, and, among those including
turbulent mixing, T6.0 and T6.09, are ruled out by our observations. 
All such models predict that in MS stars Li
should be either more abundant or the same as in subgiant stars.
The only model that predicts a A(Li) pattern which is qualitatively
similar to that observed, is the T6.25 model. For this model there is
a trend of decreasing A(Li) with decreasing \teff and at
the cool side MSs show less Li than SGs.
However, the model fails quantitatively because A(Li) of
the warmest stars is about 0.05 dex lower than what
is observed. The slope of A(Li) with \teff is not
perfectly reproduced. 
Models that include atomic diffusion and tachocline 
mixing \citep{pia08} do not seem to reproduce our observations, since
they provide a constant A(Li) up to 5500\,K.
The sophisticated models that, besides diffusion
and rotation, also take into account the effect
of internal gravity waves \citep{tac04},
seem to accurately predict the A(Li) 
pattern in solar-type stars, at solar metallicity \citep{cat05}.
However, Li isochrones have not yet been computed for
Population II stars.
Our observations call for new investigations into the stellar
physics, including gravity waves, atomic diffusion, winds and
turbulent mixing. The Li abundance pattern uncovered by our
observations has not been observed in field stars and opens up the
possibility that it may be peculiar to globular clusters, or, perhaps,
to NGC 6397. The cosmological lithium problem still awaits a solution. 

\begin{figure}
\centering
\includegraphics[width=9.cm,angle=0]{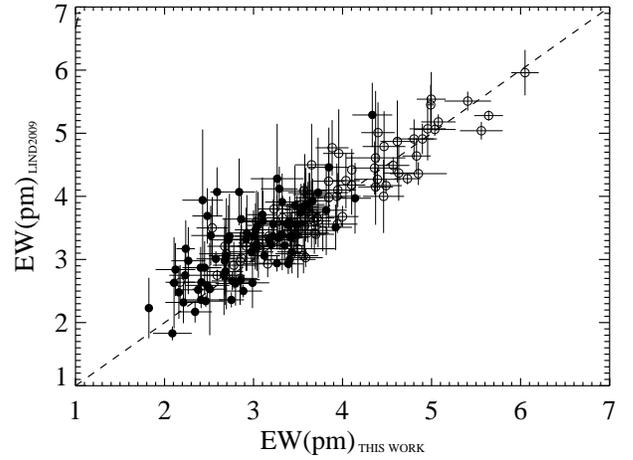}
\caption{\footnotesize{Comparison between the equivalent widths
derived in this work and those provided by Lind et al. (2009). 
Filled circles and open circles correspond to dwarf and subgiant
stars, respectively.}} 
\label{figew1d}
\end{figure}

\begin{figure}
\centering
\includegraphics[width=9.cm,angle=0]{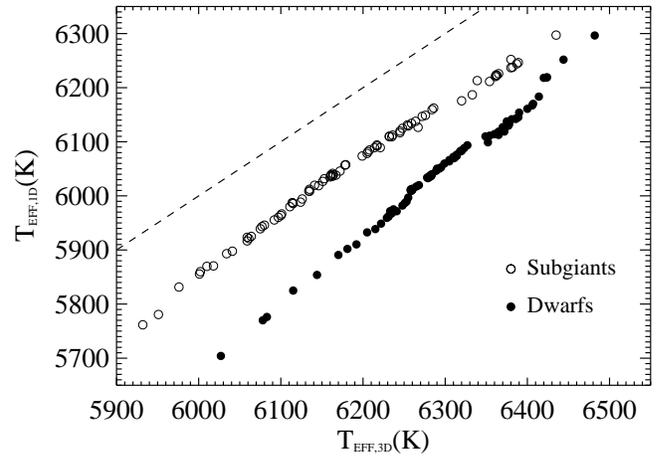}
\caption{\footnotesize{Comparison between 3D and 1D effective
temperatures of the observed stars. Filled circles and open
circles correspond to dwarf and subgiant stars, respectively. The
dashed line shows the one-to-one relationship.}}
\label{figteff1d}
\end{figure}

Our results indicate a decrease of Li abundance along the subgiant
branch, as the stars become cooler and slightly more luminous. 
This is at variance with what was found by \citet{kor07,kor06} and
\citet{lin09}, who find, instead, an {\em increase} in A(Li) in
the same region of the colour-magnitude diagram. We note that the
latter authors used our own data, as retrieved from the ESO
archive. The difference is mainly in the different \teff scales
used by the different investigations.  
\citet{lin09} also estimate slightly different EWs for our sample. 
The difference between their and our weighted mean EWs is 
$-0.08\pm0.02$~pm and $-0.08\pm0.03$~pm for SG and MS stars,
respectively (see also Fig.~\ref{figew1d} online)
The difference is smaller than the mean error in the EW measurements 
($\sim 0.2$~pm in this work and $\sim 0.35-0.4$~pm in
\citealt{lin09}), suggesting that the two sets of measurements are
fully consistent.  
To verify that the differences in EWs are irrelevant to our
conclusions we adopted the Lind et al. EWs and our \teff to
compute A(Li): our main conclusions are unchanged. 
This reinforces our claim that the difference lies in 
the \teff scale.
The difference in A(Li) that \citet{kor06} find between
turn-off (TO) and SG stars is driven by the very low \teff they find
at the TO. This is inconsistent with our H$\alpha$ fitting. Our stars
are cooler than the TO but we find higher \teff than the TO
stars in \citet{kor06}. 
We also determined 1D \teff using H$\alpha$ profiles
(see Fig.~\ref{figteff1d} online). 3D and 1D \teffo, Li
abundances and EWs of the stars in our sample are given in 
the Table~2 online. 
We compare these \teff with the colour temperatures derived
from our $B-V$ photometry and the colour calibration, based on the
infrared flux method (IRFM)
from \citet{irfm}. Adopting a mean reddening for the cluster of
$E(B-V)$=0.186 \citep{g03}, we find that for our sample of MS stars
the mean IRFM effective temperature is 6262~K, to be compared with
6047~K and 6296~K of our 1D and 3D H$\alpha$ temperatures,
respectively. The temperature spread, using both 1D and 3D H$\alpha$
fitting, is also considerably larger, by a factor of two.
That IRFM provides higher \teff than 1D H$\alpha$ is well established
\citep{irfm}. 
We repeated the analysis also with 1D model atmospheres, 
and the results are qualitatively similar: higher A(Li) for SG
stars and decreasing A(Li) for decreasing \teffo. 
The first result is very robust, since it can be deduced
directly from the distribution of Li EWs. 
The second relies on our ability to model stellar
atmospheres. To the extent that our 3D hydrodynamical models are 
a good description of a stellar atmosphere, the
second result is robust as well.
The issue of the behaviour of A(Li) with \teff
ultimately depends on the \teff scale adopted. This could be
solved if we had a direct measure of the angular diameters of
metal-poor MSs and SGs. This is probably beyond the reach 
of present-day interferometers.

NGC 6397 appears to have a higher Li content than field stars of the
same metallicity. This needs to be confirmed by 
a homogeneous analysis of field stars, with the same
models and methods. This may or may not be related to the fact 
that this cluster is nitrogen rich, compared to
field stars of the same metallicity \citep{nitrogen}.
 
\begin{acknowledgements}

We wish to thank O. Richard for providing us his
lithium depletion isochrones for different
turbulent diffusion models.
Special thanks to 
K. Lind for sending us her analysis of our data
in advance of publication. 
J. I. G. H., P. B., H.-G. L., N. B. and L. S. acknowledge
support from the EU contract MEXT-CT-2004-014265 (CIFIST).
We acknowledge use of the
supercomputing centre CINECA, which has granted us time to compute
part of the hydrodynamical models used in this investigation, through
the INAF-CINECA agreement 2006,2007.

\end{acknowledgements}

\Online

\begin{table}
\caption[]{Details of the 3D hydrodynamical model atmospheres.}  
\label{tabmod}
\centering
\begin{tabular}{lrrrr}
\hline
\hline
\noalign{\smallskip}
$\langle T_{\rm eff} \rangle$$^a$ & $\log{g}$ & [Fe/H] & Time$^b$ & Snapshots \\       
\noalign{\smallskip}
 [K] & [cgs] & [dex] & [s] &  \\       
\noalign{\smallskip}
\hline
\noalign{\smallskip}
5500 & 3.5 & -2 & 46800   & 20 \\ 
5470 & 4.0 & -2 & 33800   & 20 \\ 
5480 & 4.5 & -2 & 57000   & 20 \\ 
5860 & 3.5 & -2 & 112000  & 20 \\ 
5860 & 4.0 & -2 & 30000   & 20 \\ 
5920 & 4.5 & -2 & 24500   & 18 \\ 
6290 & 3.5 & -2 & 82800   & 18 \\ 
6280 & 4.0 & -2 & 27600   & 16 \\ 
6320 & 4.5 & -2 & 9100    & 19 \\ 
6530 & 4.0 & -2 & 49200   & 20 \\ 
6530 & 4.5 & -2 & 9100    & 19 \\ 
\noalign{\smallskip}
\hline     
\end{tabular}
\begin{list}{}{}
\item[$^{a}$] Temporal average of the emergent \teffo.
\item[$^{b}$] Total time covered by the temporal evolution of the
photospheric flow sampled at equal intervals in time named as
snapshots.
\end{list}
\end{table}

\longtab{2}{
\begin{longtable}{lrrrrrrrrr} 
\caption[]{Photometric data of the dwarf and subgiant stars of the
globular cluster NGC 6397. We also provide the signal-to-noise of the
spectra, the 3D and 1D H$\alpha$-based effective temperatures, 3D Li
abundances, and the equivalent widths and errors:
$^{\rm a}\:$Error of the equivalent width measurements estimated 
from a fitting routine that uses as free parameters 
the velocity shift, the continuum location, and the equivalent width
of the Li line.
$^{\rm b}\:$Error of the equivalent width associated to the
signal-to-noise ratio and the wavelength dispersion of the spectra,
derived using Cayrel's formula \citep{cayrel88}.
}\\  
\noalign{\smallskip}
\noalign{\smallskip}
\noalign{\smallskip}
\hline
\hline
\noalign{\smallskip}
Star Name & $V$ & $B-V$ & SNR & $T_{\rm eff,3D}$ & $T_{\rm eff,1D}$ & EW & $\delta$EW$^{a}$ & $\delta$EW$^{b}$ & A(Li)$_{\rm NLTE,3D}$ \\
          &     &       &     &   (K)	   &   (K)	    & (m\AA) & (m\AA) & (m\AA) & (dex) \\
\noalign{\smallskip}
\hline
\hline
\noalign{\smallskip}
\noalign{\smallskip}
\endfirsthead
\caption{Continued.}\\ 
\noalign{\smallskip}
\noalign{\smallskip}
\noalign{\smallskip}
\hline
\hline
\noalign{\smallskip}
Star Name & $V$ & $B-V$ & SNR & $T_{\rm eff,3D}$ & $T_{\rm eff,1D}$ & EW & $\delta$EW$^{a}$ & $\delta$EW$^{b}$ & A(Li)$_{\rm NLTE,3D}$ \\
          &     &       &     &   (K)	   &   (K)	    & (m\AA) & (m\AA) & (m\AA) & (dex) \\
\noalign{\smallskip}
\hline
\hline
\noalign{\smallskip}

\noalign{\smallskip}
\endhead
\noalign{\smallskip}
\hline
\hline
\noalign{\smallskip}
\noalign{\smallskip}
\noalign{\smallskip}
\endfoot
MSS001253 & 17.48 & 0.612 &  84 & 6279 & 6032 & 32.89 & 2.06 & 3.15 & 2.34  \\
MSS001851 & 17.38 & 0.588 &  86 & 6278 & 6033 & 24.15 & 1.97 & 3.10 & 2.19  \\
MSS002016 & 17.37 & 0.576 & 107 & 6241 & 5971 & 22.32 & 1.79 & 2.48 & 2.13  \\
MSS002882 & 17.40 & 0.598 &  83 & 6299 & 6059 & 22.12 & 2.03 & 3.18 & 2.17  \\
MSS002984 & 17.40 & 0.627 &  84 & 6365 & 6118 & 24.06 & 2.05 & 3.14 & 2.25  \\
MSS003361 & 17.32 & 0.600 &  85 & 6482 & 6296 & 34.06 & 1.78 & 3.10 & 2.51  \\
MSS004052 & 17.36 & 0.586 & 115 & 6352 & 6099 & 30.99 & 1.69 & 2.31 & 2.37  \\
MSS004099 & 17.38 & 0.597 &  82 & 6265 & 6017 & 33.21 & 1.92 & 3.22 & 2.34  \\
MSS004509 & 17.50 & 0.597 &  84 & 6318 & 6081 & 31.94 & 2.17 & 3.17 & 2.36  \\
MSS004829 & 17.44 & 0.608 &  80 & 6268 & 6019 & 27.61 & 2.31 & 3.32 & 2.25  \\
MSS005245 & 17.30 & 0.596 & 115 & 6370 & 6127 & 28.38 & 1.78 & 2.31 & 2.34  \\
MSS005478 & 17.39 & 0.587 &  89 & 6229 & 5959 & 41.40 & 1.79 & 2.97 & 2.42  \\
MSS005528 & 17.34 & 0.588 &  90 & 6375 & 6137 & 25.08 & 1.94 & 2.96 & 2.28  \\
MSS005634 & 17.41 & 0.580 & 102 & 6327 & 6093 & 32.21 & 1.90 & 2.60 & 2.37  \\
MSS005657 & 17.31 & 0.584 &  92 & 6420 & 6218 & 38.46 & 2.23 & 2.87 & 2.52  \\
MSS005755 & 17.45 & 0.600 &  97 & 6170 & 5890 & 33.90 & 2.07 & 2.72 & 2.28  \\
MSS006049 & 17.43 & 0.581 &  83 & 6289 & 6046 & 21.07 & 0.60 & 3.19 & 2.14  \\
MSS006056 & 17.37 & 0.572 &  95 & 6406 & 6167 & 27.51 & 1.41 & 2.79 & 2.35  \\
MSS006236 & 17.49 & 0.590 &  90 & 6311 & 6069 & 24.79 & 2.49 & 2.96 & 2.23  \\
MSS006292 & 17.39 & 0.578 &  85 & 6354 & 6111 & 24.11 & 2.47 & 3.11 & 2.25  \\
MSS006442 & 17.46 & 0.592 &  85 & 6378 & 6133 & 27.19 & 2.16 & 3.10 & 2.32  \\
MSS006561 & 17.37 & 0.583 &  71 & 6259 & 6012 & 27.97 & 2.25 & 3.71 & 2.25  \\
MSS006632 & 17.40 & 0.606 &  78 & 6284 & 6039 & 43.34 & 2.24 & 3.41 & 2.49  \\
MSS006666 & 17.30 & 0.598 &  82 & 6258 & 6011 & 25.77 & 2.20 & 3.22 & 2.21  \\
MSS006761 & 17.48 & 0.594 & 100 & 6293 & 6050 & 36.64 & 2.01 & 2.65 & 2.41  \\
MSS006851 & 17.32 & 0.601 &  92 & 6144 & 5853 & 34.65 & 1.99 & 2.89 & 2.27  \\
MSS006924 & 17.35 & 0.578 &  86 & 6424 & 6219 & 24.63 & 2.03 & 3.09 & 2.31  \\
MSS007267 & 17.38 & 0.605 &  82 & 6386 & 6142 & 26.77 & 0.29 & 3.24 & 2.32  \\
MSS007413 & 17.37 & 0.573 &  95 & 6205 & 5932 & 29.16 & 1.89 & 2.80 & 2.23  \\
MSS007626 & 17.42 & 0.582 &  89 & 6314 & 6073 & 32.63 & 2.07 & 2.97 & 2.37  \\
MSS007823 & 17.39 & 0.597 &  96 & 6365 & 6112 & 25.90 & 0.75 & 2.76 & 2.29  \\
MSS007830 & 17.34 & 0.587 &  98 & 6390 & 6153 & 21.61 & 0.21 & 2.71 & 2.22  \\
MSS007874 & 17.41 & 0.591 &  70 & 6310 & 6072 & 32.99 & 2.52 & 3.79 & 2.37  \\
MSS008138 & 17.44 & 0.590 & 106 & 6314 & 6076 & 24.81 & 1.69 & 2.50 & 2.23  \\
MSS008427 & 17.41 & 0.582 &  80 & 6305 & 6063 & 33.80 & 1.96 & 3.31 & 2.38  \\
MSS013089 & 17.42 & 0.597 &  84 & 6281 & 6034 & 20.88 & 2.17 & 3.16 & 2.13  \\
MSS013492 & 17.48 & 0.618 &  95 & 6231 & 5961 & 29.25 & 1.97 & 2.78 & 2.25  \\
MSS013620 & 17.31 & 0.601 &  97 & 6215 & 5938 & 31.20 & 1.77 & 2.73 & 2.27  \\
MSS013907 & 17.44 & 0.600 &  90 & 6231 & 5962 & 30.52 & 1.22 & 2.93 & 2.27  \\
MSS014243 & 17.38 & 0.606 &  86 & 6295 & 6052 & 32.17 & 2.40 & 3.07 & 2.34  \\
MSS015161 & 17.50 & 0.609 &  79 & 6181 & 5901 & 31.05 & 2.43 & 3.37 & 2.24  \\
MSS015364 & 17.36 & 0.609 &  80 & 6444 & 6251 & 18.24 & 0.05 & 3.32 & 2.18  \\
MSS016174 & 17.36 & 0.589 &  80 & 6423 & 6218 & 31.74 & 2.09 & 3.32 & 2.43  \\
MSS016301 & 17.32 & 0.621 &  94 & 6259 & 6009 & 30.20 & 1.96 & 2.81 & 2.29  \\
MSS016358 & 17.47 & 0.602 &  76 & 6407 & 6170 & 30.22 & 2.33 & 3.47 & 2.39  \\
MSS016481 & 17.40 & 0.594 &  48 & 6078 & 5770 & 37.25 & 3.35 & 5.44 & 2.26  \\
MSS016718 & 17.41 & 0.584 &  86 & 6258 & 6010 & 21.23 & 2.23 & 3.09 & 2.12  \\
MSS016850 & 17.44 & 0.588 &  84 & 6372 & 6118 & 34.32 & 1.65 & 3.16 & 2.43  \\
MSS017006 & 17.40 & 0.590 &  70 & 6305 & 6066 & 29.76 & 2.47 & 3.76 & 2.31  \\
MSS017148 & 17.37 & 0.599 & 102 & 6323 & 6088 & 30.38 & 2.10 & 2.61 & 2.34  \\
MSS017355 & 17.33 & 0.588 & 125 & 6281 & 6036 & 33.90 & 1.67 & 2.12 & 2.36  \\
MSS018410 & 17.32 & 0.591 &  99 & 6320 & 6083 & 33.49 & 2.34 & 2.68 & 2.38  \\
MSS019380 & 17.42 & 0.587 &  77 & 6253 & 5989 & 33.22 & 2.03 & 3.45 & 2.33  \\
MSS019711 & 17.40 & 0.588 &  82 & 6255 & 5996 & 30.00 & 2.14 & 3.22 & 2.28  \\
MSS019748 & 17.50 & 0.598 &  89 & 6386 & 6142 & 35.23 & 2.32 & 2.98 & 2.46  \\
MSS019966 & 17.48 & 0.592 &  77 & 6360 & 6114 & 28.85 & 2.10 & 3.44 & 2.34  \\
MSS020053 & 17.48 & 0.617 &  63 & 6377 & 6129 & 27.29 & 2.43 & 4.21 & 2.32  \\
MSS020239 & 17.44 & 0.590 &  62 & 6115 & 5824 & 32.65 & 2.96 & 4.25 & 2.22  \\
MSS020289 & 17.45 & 0.603 & 100 & 6349 & 6109 & 35.75 & 2.32 & 2.66 & 2.44  \\
MSS020400 & 17.42 & 0.604 & 116 & 6372 & 6126 & 26.84 & 1.98 & 2.29 & 2.31  \\
MSS020449 & 17.42 & 0.590 &  69 & 6248 & 5982 & 28.55 & 2.15 & 3.84 & 2.25  \\
MSS020824 & 17.32 & 0.574 &  89 & 6381 & 6141 & 26.87 & 1.87 & 2.98 & 2.32  \\
MSS020882 & 17.48 & 0.582 &  65 & 6233 & 5971 & 26.68 & 2.35 & 4.08 & 2.21  \\
MSS024313 & 17.38 & 0.597 &  78 & 6389 & 6145 & 35.88 & 2.10 & 3.39 & 2.47  \\
MSS024953 & 17.46 & 0.603 &  93 & 6027 & 5704 & 24.28 & 2.28 & 2.86 & 2.01  \\
MSS025117 & 17.31 & 0.591 &  91 & 6260 & 6013 & 23.43 & 1.90 & 2.90 & 2.17  \\
MSS025164 & 17.43 & 0.596 &  86 & 6251 & 5986 & 22.38 & 1.04 & 3.10 & 2.14  \\
MSS025647 & 17.31 & 0.597 & 121 & 6400 & 6160 & 34.88 & 1.68 & 2.20 & 2.46  \\
MSS026667 & 17.45 & 0.577 &  90 & 6283 & 6036 & 38.17 & 2.31 & 2.94 & 2.42  \\
MSS029201 & 17.41 & 0.627 &  80 & 6083 & 5776 & 22.67 & 1.90 & 3.30 & 2.02  \\
MSS029608 & 17.48 & 0.585 &  80 & 6192 & 5910 & 32.80 & 2.68 & 3.31 & 2.28  \\
MSS036470 & 17.50 & 0.610 &  77 & 6319 & 6083 & 34.19 & 2.20 & 3.44 & 2.39  \\
MSS036731 & 17.35 & 0.595 & 101 & 6414 & 6183 & 24.50 & 1.91 & 2.63 & 2.30  \\
MSS037695 & 17.40 & 0.571 &  66 & 6260 & 6010 & 25.19 & 2.19 & 4.01 & 2.20  \\
MSS037993 & 17.45 & 0.601 &  95 & 6290 & 6050 & 39.23 & 1.84 & 2.79 & 2.44  \\
MSS038318 & 17.36 & 0.584 & 100 & 6237 & 5975 & 29.88 & 1.88 & 2.64 & 2.27  \\
MSS044623 & 17.40 & 0.587 &  89 & 6283 & 6039 & 23.78 & 1.88 & 2.97 & 2.19  \\
MSS047718 & 17.48 & 0.592 &  69 & 6235 & 5967 & 34.07 & 2.90 & 3.83 & 2.33  \\
MSS049487 & 17.43 & 0.598 &  71 & 6222 & 5948 & 28.61 & 2.11 & 3.73 & 2.23  \\
SGB001167 & 15.94 & 0.587 & 108 & 6339 & 6213 & 25.91 & 1.73 & 2.46 & 2.29  \\
SGB001953 & 15.96 & 0.604 &  86 & 6361 & 6220 & 30.33 & 2.06 & 3.09 & 2.39  \\
SGB002243 & 15.94 & 0.602 &  91 & 6286 & 6162 & 37.28 & 0.19 & 2.90 & 2.43  \\
SGB002302 & 15.89 & 0.614 & 120 & 6236 & 6112 & 41.04 & 0.08 & 2.22 & 2.44  \\
SGB002675 & 15.88 & 0.603 & 109 & 6232 & 6109 & 36.24 & 1.53 & 2.44 & 2.38  \\
SGB002902 & 15.95 & 0.578 & 102 & 6205 & 6078 & 29.19 & 1.79 & 2.61 & 2.25  \\
SGB002930 & 15.95 & 0.584 & 111 & 6126 & 5994 & 36.83 & 1.64 & 2.39 & 2.31  \\
SGB003140 & 15.95 & 0.580 &  77 & 6100 & 5963 & 31.59 & 2.73 & 3.42 & 2.21  \\
SGB003332 & 15.88 & 0.608 &  69 & 6161 & 6038 & 43.92 & 2.64 & 3.82 & 2.42  \\
SGB003371 & 15.96 & 0.579 &  94 & 6380 & 6236 & 36.79 & 1.91 & 2.82 & 2.50  \\
SGB003553 & 15.86 & 0.621 &  82 & 6059 & 5916 & 48.03 & 2.25 & 3.22 & 2.39  \\
SGB003556 & 15.90 & 0.615 &  68 & 6161 & 6036 & 39.42 & 2.25 & 3.88 & 2.37  \\
SGB003678 & 15.91 & 0.589 & 117 & 6164 & 6038 & 35.58 & 1.71 & 2.27 & 2.32  \\
SGB003852 & 15.95 & 0.585 &  47 & 6018 & 5870 & 34.99 & 4.16 & 5.61 & 2.20  \\
SGB003854 & 15.89 & 0.603 & 104 & 6362 & 6223 & 34.34 & 1.62 & 2.54 & 2.45  \\
SGB003930 & 15.95 & 0.582 & 106 & 6267 & 6126 & 30.19 & 1.88 & 2.51 & 2.31  \\
SGB004063 & 15.84 & 0.629 &  68 & 6114 & 5986 & 50.36 & 2.35 & 3.90 & 2.46  \\
SGB004228 & 15.87 & 0.613 &  98 & 6111 & 5980 & 45.68 & 1.82 & 2.72 & 2.41  \\
SGB004239 & 15.84 & 0.596 & 118 & 6041 & 5897 & 35.74 & 1.84 & 2.24 & 2.23  \\
SGB004288 & 15.93 & 0.584 &  65 & 6435 & 6297 & 34.79 & 2.55 & 4.05 & 2.51  \\
SGB004474 & 15.86 & 0.603 & 110 & 6134 & 6009 & 38.83 & 1.85 & 2.40 & 2.34  \\
SGB004549 & 15.94 & 0.588 &  90 & 6255 & 6129 & 26.77 & 1.97 & 2.96 & 2.25  \\
SGB004699 & 15.83 & 0.613 &  95 & 6245 & 6119 & 36.03 & 2.03 & 2.80 & 2.39  \\
SGB004904 & 15.84 & 0.598 &  68 & 6254 & 6131 & 48.52 & 3.19 & 3.86 & 2.55  \\
SGB005126 & 15.88 & 0.602 &  85 & 6151 & 6026 & 49.90 & 0.35 & 3.12 & 2.48  \\
SGB005198 & 15.89 & 0.596 &  75 & 6097 & 5960 & 44.66 & 2.40 & 3.52 & 2.38  \\
SGB005333 & 15.88 & 0.608 & 106 & 6163 & 6036 & 34.28 & 1.58 & 2.50 & 2.30  \\
SGB005417 & 15.94 & 0.572 &  90 & 6259 & 6134 & 32.01 & 1.78 & 2.96 & 2.34  \\
SGB005556 & 15.82 & 0.620 & 110 & 6114 & 5987 & 56.41 & 1.61 & 2.41 & 2.52  \\
SGB005765 & 15.84 & 0.600 &  99 & 6199 & 6073 & 21.75 & 1.65 & 2.66 & 2.11  \\
SGB005947 & 15.92 & 0.588 &  95 & 6217 & 6092 & 36.11 & 1.86 & 2.78 & 2.37  \\
SGB006102 & 15.86 & 0.582 &  70 & 6363 & 6222 & 26.95 & 3.05 & 3.78 & 2.33  \\
SGB006281 & 15.89 & 0.619 & 122 & 6161 & 6037 & 25.36 & 1.77 & 2.17 & 2.15  \\
SGB006305 & 15.86 & 0.600 &  96 & 6232 & 6110 & 39.47 & 2.03 & 2.78 & 2.42  \\
SGB006463 & 15.93 & 0.594 & 141 & 6284 & 6159 & 40.38 & 1.55 & 1.88 & 2.47  \\
SGB006585 & 15.97 & 0.577 & 109 & 6382 & 6237 & 28.64 & 1.65 & 2.44 & 2.37  \\
SGB006625 & 15.84 & 0.603 &  88 & 6135 & 6012 & 49.56 & 2.04 & 3.02 & 2.47  \\
SGB006673 & 15.96 & 0.586 &  98 & 6167 & 6038 & 31.72 & 1.81 & 2.72 & 2.26  \\
SGB007322 & 15.95 & 0.579 & 108 & 6221 & 6089 & 34.99 & 1.91 & 2.45 & 2.35  \\
SGB007495 & 15.84 & 0.591 &  88 & 6059 & 5923 & 43.71 & 1.96 & 3.01 & 2.34  \\
SGB007501 & 15.86 & 0.579 &  86 & 6135 & 6008 & 33.25 & 2.17 & 3.10 & 2.26  \\
SGB007624 & 15.80 & 0.627 &  64 & 5932 & 5761 & 55.59 & 2.40 & 4.13 & 2.37  \\
SGB007674 & 15.89 & 0.594 &  75 & 6115 & 5985 & 43.60 & 2.50 & 3.54 & 2.38  \\
SGB008019 & 15.87 & 0.575 & 113 & 6206 & 6081 & 41.02 & 1.54 & 2.34 & 2.42  \\
SGB008043 & 15.92 & 0.599 & 117 & 6276 & 6148 & 34.27 & 1.81 & 2.26 & 2.38  \\
SGB008308 & 15.97 & 0.597 & 101 & 6333 & 6186 & 35.85 & 1.94 & 2.62 & 2.45  \\
SGB008491 & 15.90 & 0.577 &  92 & 6320 & 6175 & 36.52 & 1.88 & 2.87 & 2.45  \\
SGB008808 & 15.91 & 0.572 & 100 & 6217 & 6094 & 34.55 & 1.61 & 2.65 & 2.34  \\
SGB013359 & 15.78 & 0.629 & 129 & 6077 & 5943 & 49.96 & 1.63 & 2.06 & 2.43  \\
SGB014992 & 15.95 & 0.571 & 100 & 6064 & 5925 & 35.61 & 1.66 & 2.65 & 2.24  \\
SGB015032 & 15.93 & 0.585 &  66 & 6061 & 5921 & 35.64 & 2.56 & 3.99 & 2.24  \\
SGB015177 & 15.93 & 0.579 & 102 & 6380 & 6252 & 30.54 & 1.56 & 2.60 & 2.40  \\
SGB015392 & 15.96 & 0.586 &  85 & 6354 & 6211 & 37.14 & 1.95 & 3.13 & 2.48  \\
SGB015418 & 15.82 & 0.614 &  98 & 5951 & 5780 & 34.30 & 1.68 & 2.70 & 2.14  \\
SGB015847 & 15.81 & 0.610 & 131 & 6101 & 5966 & 39.99 & 1.80 & 2.03 & 2.33  \\
SGB016013 & 15.82 & 0.590 &  75 & 6080 & 5945 & 44.61 & 2.31 & 3.55 & 2.37  \\
SGB016363 & 15.95 & 0.577 &  71 & 6092 & 5955 & 32.63 & 2.35 & 3.74 & 2.22  \\
SGB016701 & 15.83 & 0.615 & 109 & 6141 & 6019 & 47.29 & 1.45 & 2.44 & 2.45  \\
SGB016858 & 15.98 & 0.572 &  82 & 6387 & 6244 & 28.49 & 1.94 & 3.22 & 2.37  \\
SGB016871 & 15.73 & 0.628 & 139 & 6010 & 5869 & 60.50 & 1.54 & 1.91 & 2.48  \\
SGB016936 & 15.91 & 0.602 & 117 & 6172 & 6045 & 37.78 & 1.69 & 2.26 & 2.35  \\
SGB017040 & 15.74 & 0.628 & 114 & 6002 & 5859 & 48.31 & 1.58 & 2.33 & 2.35  \\
SGB017100 & 15.83 & 0.612 &  51 & 5703 & 5506 & 43.71 & 3.35 & 5.16 & 2.06  \\
SGB017116 & 15.96 & 0.589 &  82 & 6124 & 5988 & 35.34 & 1.84 & 3.24 & 2.29  \\
SGB018051 & 15.93 & 0.575 &  88 & 6264 & 6137 & 35.09 & 1.96 & 3.02 & 2.39  \\
SGB018096 & 15.92 & 0.585 &  96 & 6214 & 6089 & 37.00 & 2.04 & 2.75 & 2.38  \\
SGB018128 & 15.84 & 0.621 &  90 & 6153 & 6031 & 46.31 & 1.76 & 2.95 & 2.45  \\
SGB018930 & 15.88 & 0.606 &  92 & 6179 & 6057 & 36.31 & 1.86 & 2.87 & 2.34  \\
SGB019686 & 15.89 & 0.591 & 131 & 6365 & 6225 & 34.16 & 1.54 & 2.03 & 2.45  \\
SGB019890 & 15.95 & 0.607 & 106 & 6146 & 6018 & 44.88 & 1.78 & 2.49 & 2.42  \\
SGB020001 & 15.76 & 0.613 & 124 & 5976 & 5831 & 57.08 & 1.46 & 2.15 & 2.42  \\
SGB020304 & 15.84 & 0.580 & 100 & 6248 & 6124 & 39.57 & 1.75 & 2.66 & 2.44  \\
SGB024422 & 15.81 & 0.608 & 101 & 6075 & 5939 & 48.97 & 1.63 & 2.62 & 2.41  \\
SGB024914 & 15.95 & 0.602 &  72 & 6235 & 6109 & 27.78 & 2.35 & 3.68 & 2.25  \\
SGB025290 & 15.92 & 0.582 & 103 & 6272 & 6146 & 38.43 & 1.81 & 2.57 & 2.44  \\
SGB025764 & 15.96 & 0.584 &  89 & 6245 & 6116 & 28.50 & 1.60 & 2.99 & 2.27  \\
SGB026642 & 15.81 & 0.606 &  62 & 6160 & 6035 & 39.36 & 2.67 & 4.24 & 2.37  \\
SGB029317 & 15.79 & 0.603 &  96 & 6001 & 5855 & 50.76 & 1.93 & 2.76 & 2.38  \\
SGB029417 & 15.77 & 0.619 &  82 & 6034 & 5893 & 38.45 & 1.87 & 3.23 & 2.26  \\
SGB030350 & 15.76 & 0.619 &  62 & 6178 & 6056 & 54.06 & 2.62 & 4.29 & 2.55  \\
SGB030403 & 15.91 & 0.613 & 100 & 6389 & 6246 & 36.51 & 1.91 & 2.64 & 2.50  \\
SGB031394 & 15.96 & 0.572 &  92 & 6362 & 6222 & 32.30 & 1.81 & 2.88 & 2.42  \\
SGB032079 & 15.78 & 0.625 &  89 & 6163 & 6041 & 43.98 & 1.88 & 2.97 & 2.43  \\
SGB036901 & 15.95 & 0.606 &  82 & 6208 & 6084 & 46.15 & 1.91 & 3.21 & 2.48  \\
\label{BIGtable}  
\end{longtable}
}
 

\begin{thebibliography}{}

\bibitem[Aller \& Chapman(1960)]{aac60} 
Aller, L.~H., \& Chapman, S.\ 1960, \apj, 132, 461 

\bibitem[Asplund et al. (2006)]{asp06}
Asplund, M., Lambert, D. L., Nissen, P. E., Primas, F., \& Smith, V.
V. 2006, \apj, 644, 229  

\bibitem[Barklem et al.(2000)]{bar00} 
Barklem, P.~S., Piskunov, N., \& O'Mara, B.~J.\ 2000, \aap, 363, 1091 

\bibitem[Behara et al.(2009)]{beh09} 
Behara, N.~T., Ludwig, H.-G., Steffen, M., \& Bonifacio, P.\ 2009,
American Institute of Physics Conference Series, 1094, 784 

\bibitem[Boesgaard et al.(1998)]{boe98} Boesgaard, A.~M., 
Deliyannis, C.~P., Stephens, A., \& King, J.~R.\ 1998, \apj, 493, 206 

\bibitem[Bonifacio et al.(2002)]{bon02}
Bonifacio, P., et al.\ 2002, \aap, 390, 91 

\bibitem[Bonifacio et al.(2007a)]{bon07} 
Bonifacio, P., et al.\ 2007a, \aap, 462, 851 

\bibitem[Bonifacio et al.(2007b)]{bonifacio} Bonifacio, P., 
et al.\ 2007b, \aap, 470, 153 


\bibitem[Carlsson et al.(1994)]{carlsson} Carlsson, M., 
Rutten, R.~J., Bruls, J.~H.~M.~J., \& Shchukina, N.~G.\ 1994, \aap, 288, 860 

\bibitem[Cayrel(1988)]{cayrel88} Cayrel, R.\ 1988, IAU
Symp.~132: The Impact of Very High S/N Spectroscopy on Stellar Physics,
132, 345

\bibitem[Cayrel et al.(2007)]{cay07} 
Cayrel, R., et al.\ 2007, \aap, 473, L37 

\bibitem[Charbonnel \& Primas(2005)]{cap05} 
Charbonnel, C., \& Primas, F.\ 2005, \aap, 442, 961 

\bibitem[Charbonnel \& Talon(2005)]{cat05} 
Charbonnel, C., \& Talon, S.\ 2005, Science, 309, 2189 

\bibitem[Cyburt et al.(2008)]{cyburt} Cyburt, R.~H., Fields, 
B.~D., 
\& Olive, K.~A.\ 2008, Journal of Cosmology and Astro-Particle Physics, 11, 12 

\bibitem[Freytag et al.(2002)]{fre02} 
Freytag, B., Steffen, M., \& Dorch, B.\ 2002, Astronomische
Nachrichten, 323, 213  

\bibitem[Gonz{\'a}lez Hern{\'a}ndez \& Bonifacio(2009)]{irfm} 
Gonz{\'a}lez Hern{\'a}ndez, J.~I., \& Bonifacio, P.\ 2009, \aap, 497, 497 

\bibitem[Gonz{\'a}lez Hern{\'a}ndez et al.(2008)]{gon08} 
Gonz{\'a}lez Hern{\'a}ndez, J.~I., et al.\ 2008, \aap, 480, 233 

\bibitem[Gratton et al.(2001)]{gratton} Gratton, R.~G., et al.\ 2001, \aap, 369, 87 

\bibitem[Gratton et al.(2003)]{g03} Gratton, R.~G., 
Bragaglia, A., Carretta, E., Clementini, G., 
Desidera, S., Grundahl, F., \& Lucatello, S.\ 2003, \aap, 408, 529 

\bibitem[Griem(1960)]{gri60} 
Griem, H.~R.\ 1960, \apj, 132, 883 

\bibitem[Hisano et al.(2009)]{hisano} 
Hisano, J., Kawasaki, M., Kohri, K., \& Nakayama, K.\ 2009, \prd, 79,
063514  

\bibitem[Jedamzik(2004)]{jed04} 
Jedamzik, K.\ 2004, \prd, 70, 083510 

\bibitem[Jedamzik(2006)]{jed06} 
Jedamzik, K.\ 2006, \prd, 74, 103509 

\bibitem[Jittoh et al.(2008)]{jittoh} 
Jittoh, T., Kohri, K., Koike, M., Sato, J., Shimomura, T., \&
Yamanaka, M.\ 2008, \prd, 78, 055007  

\bibitem[Korn et al.(2006)]{kor06} 
Korn, A.~J., Grundahl, F., Richard, O., Barklem, P.~S., Mashonkina,
L., Collet, R., Piskunov, N., \& Gustafsson, B.\ 2006, \nat, 442, 657

\bibitem[Korn et al.(2007)]{kor07} 
Korn, A.~J., Grundahl, F., Richard, O., Mashonkina, L., Barklem,
P.~S., Collet, R., Gustafsson, B., \& Piskunov, N.\ 2007, \apj, 671,
402  

 
\bibitem[Lind et al.(2009)]{lin09} 
Lind, K. et al. \ 2009, \aap, in press

\bibitem[Ludwig et al.(2009)]{lud09} 
Ludwig, H.-G., Behara, N.~T., Steffen, M., \& Bonifacio, P.\ 2009,
\aap, 502, L1  

\bibitem[Mel{\'e}ndez \& Ram{\'{\i}}rez(2004)]{mel04} 
Mel{\'e}ndez, J., \& Ram{\'{\i}}rez, I.\ 2004, \apjl, 615, L33 

\bibitem[Michaud et al.(1984)]{mic84} 
Michaud, G., Fontaine, G., \& Beaudet, G.\ 1984, \apj, 282, 206 
\bibitem[Molaro 
\& Pasquini(1994)]{MP94} Molaro, P., \& Pasquini, L.\ 1994, \aap, 281, L77 

\bibitem[Pasquini 
\& Molaro(1997)]{PM97} Pasquini, L., \& Molaro, P.\ 1997, \aap, 322, 109 

\bibitem[Pasquini 
\& Molaro(1996)]{PM96} Pasquini, L., \& Molaro, P.\ 1996, \aap, 307, 761 

\bibitem[Pasquini et al.(2002)]{pas02} 
Pasquini, L., et al.\ 2002, The Messenger, 110, 1 

\bibitem[Pasquini et al.(2005)]{pasquini} Pasquini, L., Bonifacio, P., 
Molaro, P., 
Fran\c ois, P., Spite, F., Gratton, R.~G., 
Carretta, E., \& Wolff, B.\ 2005, \aap, 441, 549 

\bibitem[Pasquini et al.(2008)]{nitrogen} Pasquini, L., Ecuvillon, A., 
Bonifacio, P., \& Wolff, B.\ 2008, \aap, 489, 315 

\bibitem[Piau et al.(2006)]{pia06} 
Piau, L., Beers, T.~C., Balsara, D.~S., Sivarani, T., Truran, J.~W., 
\& Ferguson, J.~W.\ 2006, \apj, 653, 300 

\bibitem[Piau(2008)]{pia08} 
Piau, L.\ 2008, \apj, 689, 1279 

\bibitem[Richard et al.(2002)]{ric02} 
Richard, O., Michaud, G., \& Richer, J.\ 2002, \apj, 580, 1100 

\bibitem[Richard et al.(2005)]{ric05} 
Richard, O., Michaud, G., \& Richer, J.\ 2005, \apj, 619, 538 

\bibitem[Sbordone et al.(2008)]{sbor08} Sbordone, L., et al.\ 
2008, First Stars III Conference. AIP Conference 
Proceedings, Volume  990, p. 339 


\bibitem[Spergel et al.(2007)]{spe07} 
Spergel, D.~N., et al.\ 2007, \apjs, 170, 377 

\bibitem[Spite \& Spite(1982a)]{spi82a} 
Spite, M., \& Spite, F.\ 1982a, \nat, 297, 483 

\bibitem[Spite \& Spite(1982b)]{spi82b} 
Spite, F., \& Spite, M.\ 1982b, \aap, 115, 357 

\bibitem[Straniero et al.(1997)]{str97} 
Straniero, O., Chieffi, A., \& Limongi, M.\ 1997, \apj, 490, 425 

\bibitem[Talon \& Charbonnel(2004)]{tac04} 
Talon, S., \& Charbonnel, C.\ 2004, \aap, 418, 1051 

\bibitem[Th{\'e}venin et al.(2001)]{theve} Th{\'e}venin, F., 
Charbonnel, C., de Freitas Pacheco, J.~A., Idiart, T.~P., 
Jasniewicz, G., de Laverny, P., \& Plez, B.\ 2001, \aap, 373, 905 

\bibitem[Vidal et al.(1973)]{vid73} 
Vidal, C.~R., Cooper, J., \& Smith, E.~W.\ 1973, \apjs, 25, 37 

\bibitem[Wedemeyer et al.(2004)]{wed04}
Wedemeyer, S., Freytag, B., Steffen, M., Ludwig, H.-G., \& Holweger,
H.\ 2004, \aap, 414, 1121  


\end{thebibliography}
\end{document}